# Automated mapping of reaction phase diagrams in complex oxides

Jianghao Jin[1,2], Jinlin Feng[1,2], Yidong Huang[1], Sheng Meng[1]*, Miao Liu[1,2]*

Institute of Physics, Chinese Academy of Sciences, Beijing 100190, China;

Dongguan Institute of Materials Science and Technology, Dongguan, Guangdong 523808,China

* E-mail: smeng@iphy.ac.cn, mliu@iphy.ac.cn

## Abstract

Systematic, internally consistent experimental phase-diagram data are far less available for inorganic ionic compounds than for alloys, despite the central role of phase selection in the synthesis of functional oxides. Here we develop an automated laboratory platform that maps reaction phase diagrams under a fixed reaction time, combining parallel robotic synthesis with iterative phase-boundary refinement. Each experiment requires an 8 h reaction, and parallel operation of multiple furnaces under continuous 24/7 robotic scheduling completes a pseudo-binary reaction phase diagram within a few days. We demonstrate the approach on the Mg–V–O and Ba–Mn–O systems along the MgO–$V_2O_5$ and BaO–$MnO_2$ precursor sections. The Mg–V–O system reacts without measurable oxygen loss and shows diffusion-controlled phase evolution with sharp boundaries and well-defined polymorphic transformations, whereas the Ba–Mn–O system releases oxygen continuously and exhibits broad multiphase coexistence coupled to Mn valence evolution. These contrasting cases establish a protocol for constructing reaction phase diagrams and provide a basis for a standardized reaction phase-diagram database for inorganic compounds.

**Keywords:** automated laboratory; phase diagram; high-throughput experimentation; automated synthesis; solid-state reaction; reaction phase diagram; reaction kinetics; valence evolution; oxides

## 1. Introduction

Phase diagrams are widely regarded as the "maps" of physical chemistry and materials science, and they underlie virtually every aspect of materials design—from alloy selection to functional ceramics to battery cathode engineering. From a thermodynamic perspective, a phase diagram represents a macroscopic projection of the Gibbs free energy of a system as a function of temperature (T), pressure (P), and chemical composition (x) [1]. According to the Clausius–Clapeyron equation and the Gibbs phase rule ($F = C - P + 2$), phase diagrams theoretically define the forms in which substances exist under thermodynamic equilibrium conditions, and the boundaries between different phase regions are determined by the equality of chemical potentials among coexisting phases [1–2].

The metallurgical community enjoys a mature phase-diagram infrastructure of standardized databases, commercial CALPHAD packages, and decades of curated data. No equivalent, internally consistent experimental dataset exists for inorganic ionic-compound systems: an Mg–V–O or Ba–Mn–O phase diagram, for example, must still be reconstructed from scattered publications that use inconsistent precursor choices, reaction times, and characterization protocols.

The limited availability of comparable data has practical consequences: synthesis windows for functional oxides (battery cathodes, catalysts, multiferroics, superconductors) are typically identified by empirical trial and error. The challenge is not a lack of thermodynamic theory, but the lack of experimental phase-diagram data obtained under consistent precursor, time, atmosphere, and characterization conditions. Constructing such datasets manually is impractical for ternary and higher-order systems because the number of required compositions grows rapidly. Solid-state reactions introduce further constraints: precursor selection, diffusion rates, reaction time, and heating pathways can prevent the system from reaching equilibrium [3–5]. Reaction phase diagrams address this problem by recording phase formation under explicitly specified reaction conditions rather than at the equilibrium limit.

Traditional phase-diagram construction relies on extensive experimental measurements that are time-consuming and dependent on empirical judgment [6]. The CALPHAD method [7–8] predicts equilibrium phase regions through free-energy minimization but requires systematic thermodynamic data, which remain limited for many inorganic systems with multiple valence states and strong electron correlations [9–11]. First-principles DFT methods [12–14] provide ground-state convex-hull energies but face challenges in finite-temperature free-energy calculations, stringent energy-accuracy requirements, and the dimensionality of multicomponent systems [15]. Both approaches describe thermodynamic stability, whereas actual solid-state reactions proceed for finite times and depend on diffusion, precursor selection, and reaction pathways. Reaction phase diagrams measured under specified precursors, a fixed reaction time, and a defined heating pathway [16–18] therefore provide complementary information for synthesis.

Recent robotic laboratories—including the Ceder group's A-Lab [19–20], the Cooper group's mobile robotic chemist [21–22], and the Sun group's ASTRAL platform [23]—now perform high-throughput synthesis with high efficiency and reproducibility, and phase-diagram mapping guided by high-throughput experiments or by uncertainty- and active-learning-based sampling has also been reported [6,24]. These studies, however, are largely organized around the discovery of individual target compounds or around model-based sampling policies. What is still missing is a simple, model-free experimental protocol that converts robotic synthesis directly into reaction phase diagrams, with the precursor route and reaction time held explicitly fixed—an inherently multi-target, combinatorial measurement problem.

In this work, we show that an automated laboratory platform can construct reaction phase diagrams in complex inorganic systems by combining high-throughput robotic synthesis with iterative phase-boundary refinement. Applied to the $MgO$–$V_2O_5$ and $BaO$–$MnO_2$ pseudo-binary sections, the approach resolves a reaction phase diagram within a few days: each experiment requires an 8 h reaction, while multiple furnaces running in parallel under continuous 24/7 robotic operation provide the required throughput. The two sections reveal diffusion-controlled phase evolution in the Mg–V–O system and valence-coupled multiphase coexistence in the Ba–Mn–O system. The resulting protocol provides a basis for a standardized reaction phase-diagram database for inorganic compounds and for data-driven synthesis planning.

# 2. Experiments and results

## 2.1 Automated phase mapping

We have established an automated laboratory platform for the solid-state synthesis of inorganic materials. The platform adopts a modular architecture, in which the experimental workflow is divided into five functional modules—transport, automatic batching, automated mixing, heat treatment, and characterization—all coordinated through a central control system to achieve unified scheduling and process integration. The transport module is centered on an automated guided vehicle (AGV), equipped with a robotic arm and an Intel RealSense depth camera. By recognizing pre-set QR codes at each module, relative coordinate re-localization and correction are achieved, thereby reducing the impact of navigation errors on material transfer accuracy. The automatic batching module performs material dispensing through the coordination of a robotic arm and a Mettler Toledo automated balance. Programmable mass control is implemented based on predefined formulation parameters, minimizing fluctuations arising from manual operation. The automated mixing module consists of a mixing device, a robotic arm, and a rotary motor, enabling automated opening of sample containers, material mixing, and transfer operations. The uniformity and batch-to-batch consistency of samples are improved by standardizing mixing time and rotation speed. The heat treatment module employs a muffle furnace to execute predefined heating, dwelling, and cooling programs, ensuring temperature control accuracy and experimental consistency under isothermal reaction conditions. Subsequently, the samples are transferred to the characterization module for X-ray diffraction (XRD) analysis, enabling automated acquisition of structural information.

The platform supports continuous 24/7 operation, and several muffle furnaces are run in parallel so that different composition–temperature points are annealed simultaneously. This parallel scheduling, rather than the 8 h reaction time of an individual experiment, sets the wall-clock time required to complete a diagram. Standardized operation links each product to a defined precursor ratio and thermal history, providing a common protocol for comparing reaction phase diagrams across inorganic systems.

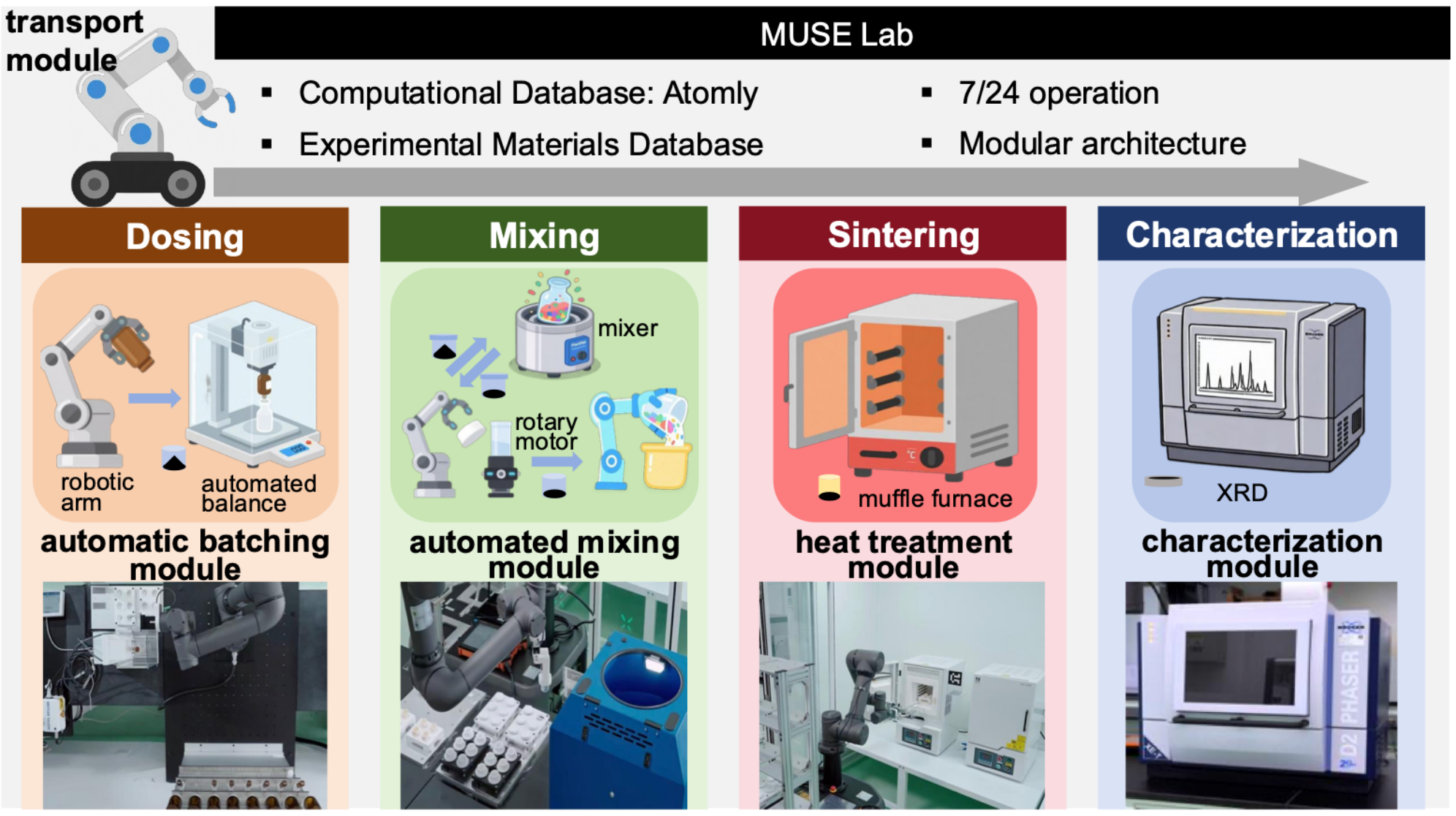

**Figure 1.** Schematic illustration and photographic overview of the automated laboratory platform (MUSE Lab) for inorganic materials synthesis.

To validate the method, we selected the Mg–V–O and Ba–Mn–O systems because they provide a controlled chemical contrast rather than two unrelated demonstrations. Along the $MgO$–$V_2O_5$ section, vanadium remains essentially pentavalent and no measurable mass loss is detected, so the reaction proceeds in an effectively closed oxygen system. Along the $BaO$–$MnO_2$ section, manganese changes valence and oxygen is released continuously, so the reaction proceeds in an open oxygen system. Both sections contain several stoichiometric compounds with distinct structure types and therefore exhibit well-defined yet non-trivial phase-evolution pathways, making them suitable for testing phase-region classification and phase-boundary localization. The two systems also differ in cation radius, coordination environment, and valence flexibility, and their compounds are relevant to energy and functional oxide materials, so the resulting diagrams are useful both as methodological benchmarks and as practical synthesis guidance.

Reaction phase diagrams were constructed by robotic synthesis along fixed precursor sections, with all samples reacted isothermally for 8 h. An initial 100 °C temperature grid was supplemented by temperatures selected from DSC/TGA, and product assemblages were identified by XRD. Temperatures near phase changes were refined iteratively at each composition; composition boundaries were assigned from the observed assemblages and phase stoichiometries. Fixing the reaction time ensures direct comparability among experiments, so the diagrams describe kinetically constrained reaction outcomes rather than thermodynamic equilibrium. Boundary uncertainty is set by the final sampling interval and the XRD detection limit. The resulting diagrams are shown in Figures 2b and 3b.

## 2.2 Phase evolution in Mg–V–O

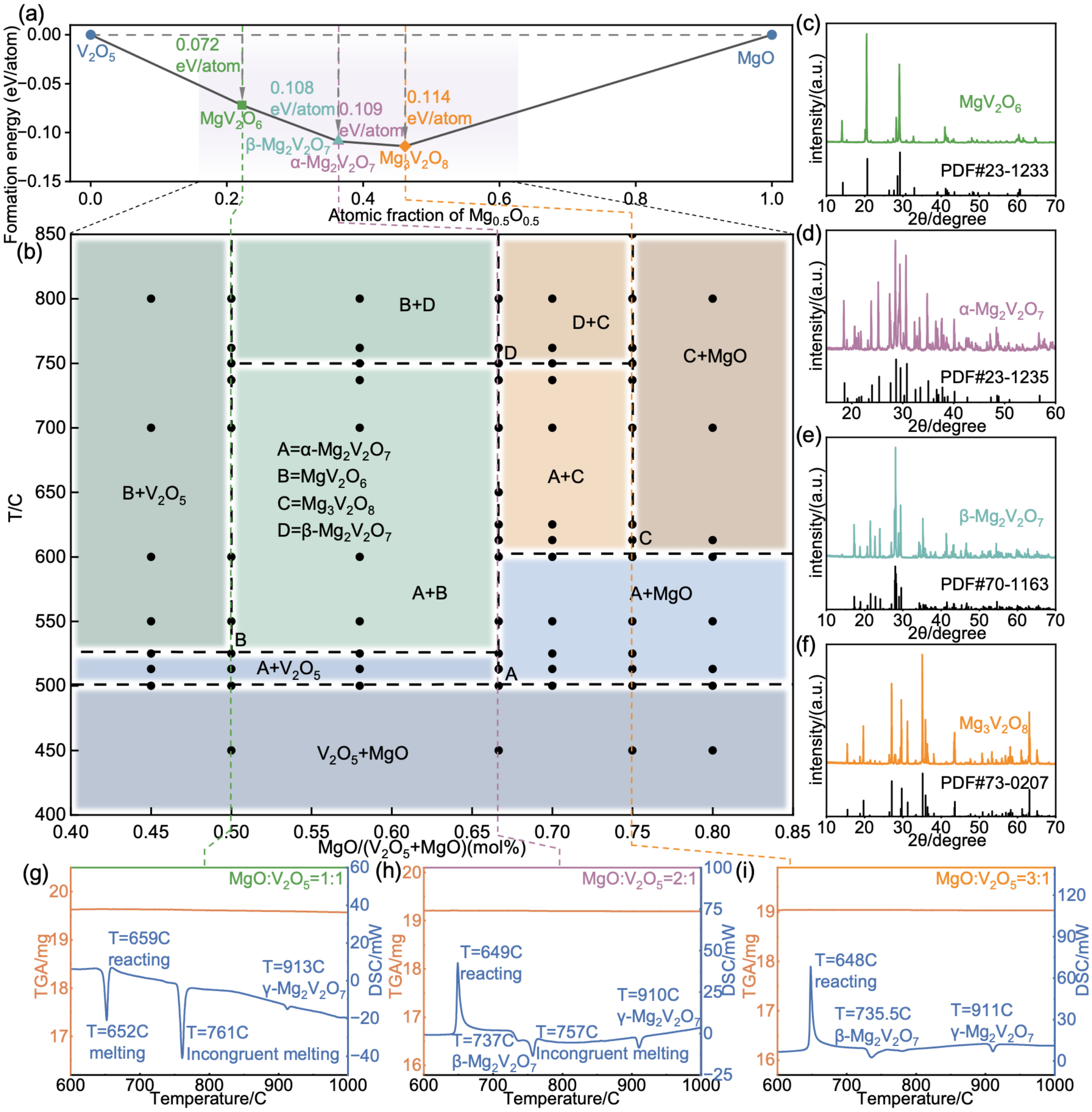

**Figure 2.** $MgO$–$V_2O_5$ system. (a) Convex hull constructed from Atomly database entries, with formation energies referenced to the oxide end members. (b) Reaction phase diagram along the $MgO$–$V_2O_5$ pseudo-binary section under a fixed 8 h reaction time: A = α-$Mg_2V_2O_7$, B = $MgV_2O_6$, C = $Mg_3V_2O_8$, D = β-$Mg_2V_2O_7$; black dots are experimental points and dashed lines are phase boundaries refined by iterative sampling. (c–f) XRD patterns of $MgV_2O_6$, α-$Mg_2V_2O_7$, β-$Mg_2V_2O_7$, and $Mg_3V_2O_8$ with the corresponding PDF cards. (g–i) TGA and DSC curves of $MgO$–$V_2O_5$ mixtures with molar ratios of 1:1, 2:1, and 3:1.

To establish a theoretical reference for the Mg–V–O system, we first analyzed ground-state stability using the Atomly database [25]. Within the compositional range defined by the MgO and $V_2O_5$ end members, the formation energies of $MgV_2O_6$ (–0.072 eV/atom), $Mg_2V_2O_7$ (–0.109 eV/atom), and $Mg_3V_2O_8$ (–0.114 eV/atom) place all three compounds on the 0 K convex hull. The α and β polymorphs of $Mg_2V_2O_7$ differ by only ≈1 meV/atom, well within the accuracy of DFT, so the 0 K hull cannot determine their relative stability. Convex hull analysis describes thermodynamic equilibrium but contains no information about diffusion, pathway dependence, or finite-time constraints. These results therefore serve as a reference for thermodynamic stability rather than as a direct predictor of the experimental reaction phase diagram.

To determine the temperature range of solid-state reactions, MgO and $V_2O_5$ were mixed at different stoichiometric ratios (1:1, 2:1, and 3:1), followed by DSC and TGA measurements. Based on the experimental results, the temperature range of 600–1000 °C was selected for subsequent analysis. The TGA curves show that the sample mass remains nearly constant throughout the heating process, with no significant weight loss observed, indicating that no gas release occurs during the reaction and that the system predominantly undergoes solid–solid or solid–liquid transformations. The DSC curves exhibit pronounced thermal signals within specific temperature ranges, indicating the onset of reactions and phase transformations. Combined with XRD characterization of the products before and after these events, a clear endothermic peak near 650 °C is assigned to melting of the $V_2O_5$-rich precursor assemblage, and the immediately following exothermic event indicates rapid reaction and formation of new phases under locally molten conditions.

With increasing MgO content, the endothermic peak and the subsequent exothermic peak gradually overlap, eventually resulting in an overall exothermic feature in the DSC curves. This phenomenon is closely related to the fact that local melting under rapid heating conditions in DSC promotes diffusion and accelerates reaction kinetics. Within this temperature range, $MgV_2O_6$ (monoclinic, space group C2/m) [26], α-$Mg_2V_2O_7$ (monoclinic, space group $P2_1/c$) [27], and $Mg_3V_2O_8$ (orthorhombic, space group Cmca) [28] are rapidly formed. When the temperature reaches approximately 735 °C, α-$Mg_2V_2O_7$ undergoes a polymorphic transformation to form β-$Mg_2V_2O_7$ (triclinic, space group $P\bar{1}$) [27]. Upon further heating to approximately 761 °C, incongruent melting occurs under conditions of relatively low MgO content. When the temperature is further increased to around 911 °C, β-$Mg_2V_2O_7$ transforms into γ-$Mg_2V_2O_7$. The γ-$Mg_2V_2O_7$ phase is stable at high temperatures but cannot be retained upon cooling to room temperature. By comparison with standard diffraction patterns, $MgV_2O_6$, α-$Mg_2V_2O_7$, β-$Mg_2V_2O_7$, and $Mg_3V_2O_8$ were successfully synthesized and identified.

The DSC/TGA results identified the thermal events used to select reaction temperatures for the automated experiments. Subsequent sampling was concentrated around reaction onset, phase formation, and polymorphic transformation regions rather than distributed uniformly across the full temperature range.

For the MgO–$V_2O_5$ system, no reaction is detected below 500 °C. Because the 8 h isothermal reaction time far exceeds the effective holding time of a DSC scan, products are already observed at ≈500 °C, below the main DSC peak temperatures.

At around 500 °C, the system preferentially forms α-$Mg_2V_2O_7$, with a small amount of unreacted precursors remaining. When the temperature increases to approximately 520 °C, $MgV_2O_6$ is detected on the $V_2O_5$-rich side, and a two-phase coexistence region of α-$Mg_2V_2O_7$ + $MgV_2O_6$ appears in the intermediate composition range. As the temperature further increases to about 600 °C, $Mg_3V_2O_8$ forms on the MgO-rich side, leading to a distinct two-phase coexistence region between α-$Mg_2V_2O_7$ and $Mg_3V_2O_8$. At approximately 750 °C in the 8 h reaction phase diagram, α-$Mg_2V_2O_7$ has completely transformed into β-$Mg_2V_2O_7$, and the phase diagram correspondingly exhibits two stable coexistence regions: β-$Mg_2V_2O_7$ + $MgV_2O_6$ and β-$Mg_2V_2O_7$ + $Mg_3V_2O_8$. The α→β boundary at ≈750 °C is consistent with the DSC events at 735–737 °C within the temperature discretization of the scan; the small offset reflects the different heating protocols and the finite interval used for boundary localization. Overall, this system exhibits a well-defined reaction onset temperature, clear and stable phase boundaries, and identifiable polymorphic transformation pathways. The reaction process is primarily governed by solid-state diffusion and can therefore be regarded as a typical diffusion-controlled reaction phase diagram.

### 2.3 Oxygen-driven reactions in Ba–Mn–O

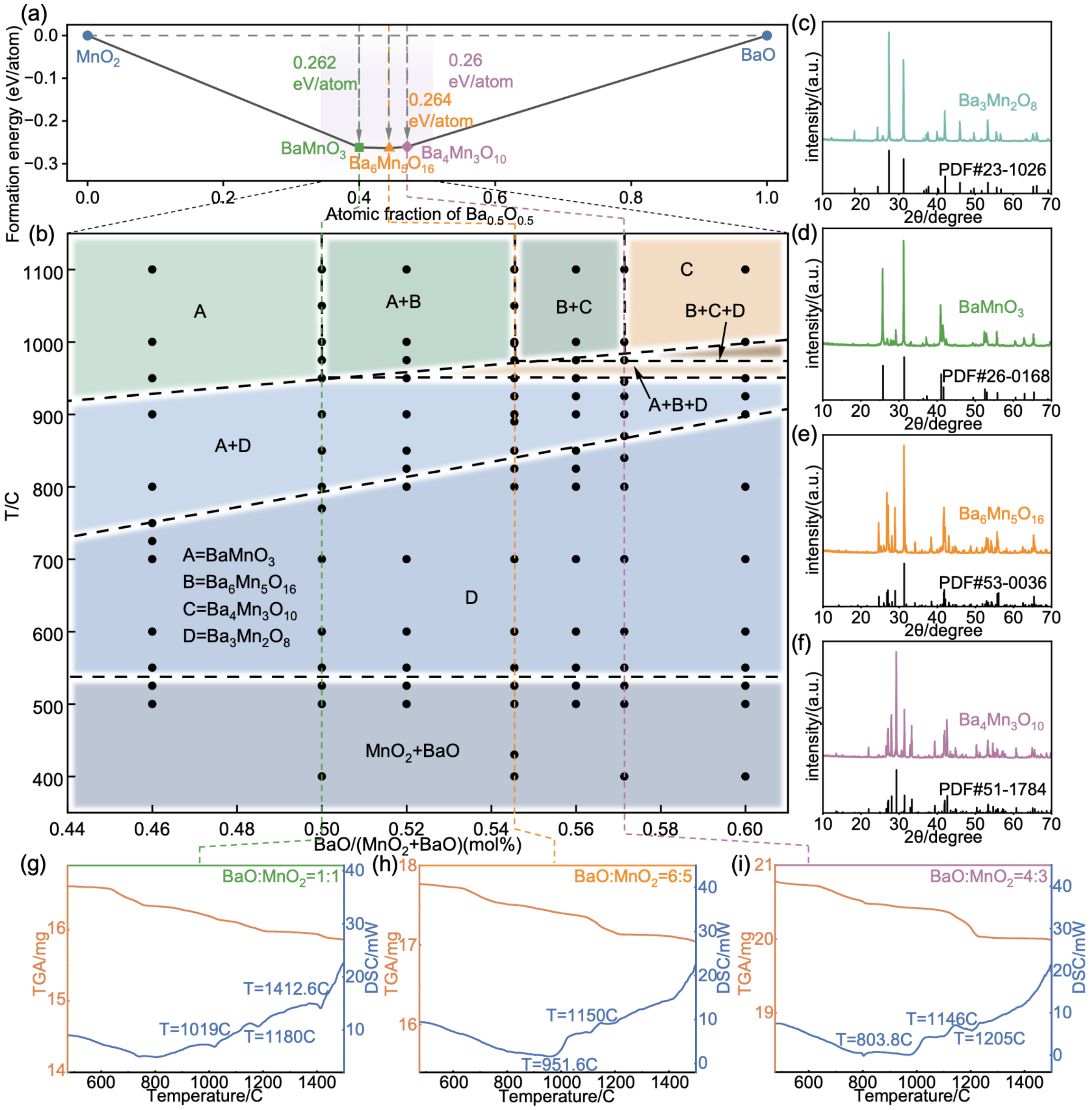

**Figure 3.** $BaO–MnO_2$ system. (a) Convex hull constructed from Atomly database entries. (b) Reaction phase diagram along the $BaO–MnO_2$ pseudo-binary section under a fixed 8 h reaction time: A = $BaMnO_3$, B = $Ba_6Mn_5O_{16}$, C = $Ba_4Mn_3O_{10}$, D = $Ba_3Mn_2O_8$; black dots are experimental points and dashed lines are phase boundaries refined by iterative sampling. (c–f) XRD patterns of $Ba_3Mn_2O_8$, $BaMnO_3$, $Ba_6Mn_5O_{16}$, and $Ba_4Mn_3O_{10}$ with the corresponding PDF cards. (g–i) TGA and DSC curves of $BaO–MnO_2$ mixtures with molar ratios of 1:1, 6:5, and 4:3.

To establish a corresponding reference for Ba–Mn–O, we evaluated the formation energies of $BaMnO_3$ (–0.262 eV/atom), $Ba_6Mn_5O_{16}$ (–0.264 eV/atom), and $Ba_4Mn_3O_{10}$ (–0.260 eV/atom) using BaO and $MnO_2$ as the end members. The three values span only 0.004 eV/atom, and all three compounds lie on or close to the 0 K convex hull. This near-degeneracy indicates strong thermodynamic competition, so modest differences in diffusion or local oxygen chemical potential may influence the products. $Ba_3Mn_2O_8$, however, is absent from these hull phases despite being the lowest-temperature product observed in the experiments. The equilibrium reference therefore does not predict the observed temperature-dependent phase evolution.

To clarify the thermal behavior of the manganese precursor, DSC/TGA measurements were first performed on $MnO_2$ alone. $MnO_2$ begins to decompose at approximately 610 °C, releasing $O_2$ and forming $Mn_2O_3$. At about 1000 °C, $Mn_2O_3$ further decomposes with oxygen release to form $Mn_3O_4$. Near 1178 °C, $Mn_3O_4$ undergoes a polymorphic transformation from a distorted tetragonal spinel to a cubic spinel [29]. This decomposition sequence provides a thermal baseline for interpreting the BaO–$MnO_2$ mixtures.

BaO and $MnO_2$ were then mixed at stoichiometric ratios of 1:1, 6:5, and 4:3 and measured by DSC/TGA from 500 to 1500 °C. The TGA curves show continuous mass loss throughout heating, indicating ongoing deoxygenation in an open oxygen system. The DSC curves contain multiple small peaks superimposed on a weak overall thermal effect, consistent with overlapping decomposition, phase-formation, and structural-transformation steps. The BaO–$MnO_2$ reaction therefore involves coupled valence evolution, solid-state diffusion, and high-temperature structural transformations.

XRD analysis of products obtained at different temperatures identified $BaMnO_3$ (hexagonal, space group $P6_3/mmc$) [30], $Ba_6Mn_5O_{16}$ (hexagonal, space group $P6_3/mmc$) [31], and $Ba_4Mn_3O_{10}$ (orthorhombic, space group Cmca) [32] by comparison with standard diffraction patterns. $Ba_3Mn_2O_8$ (trigonal, space group R-3m) [33] was also identified, although it is absent from the 0 K hull reference. These four phases provide the phase assignments used to construct Figure 3b.

The BaO–$MnO_2$ reaction phase diagram contains more multiphase fields than the MgO–$V_2O_5$ diagram, consistent with the continuous oxygen loss observed by TGA. No reaction is detected below 550 °C. $Ba_3Mn_2O_8$ is the lowest-temperature product detected by ex situ XRD. With further heating, a $Ba_3Mn_2O_8$ + $BaMnO_3$ coexistence region emerges, and the phase boundary shifts toward higher temperatures with increasing BaO content.

As the temperature continues to rise, $Ba_3Mn_2O_8$ gradually decomposes and disappears, and the system sequentially exhibits a $BaMnO_3$ single-phase region, a $BaMnO_3$ + $Ba_6Mn_5O_{16}$ coexistence region, a $Ba_6Mn_5O_{16}$ + $Ba_4Mn_3O_{10}$ coexistence region, and a $Ba_4Mn_3O_{10}$ single-phase region. In addition, three-phase fields ($BaMnO_3$ + $Ba_6Mn_5O_{16}$ + $Ba_3Mn_2O_8$ and $Ba_6Mn_5O_{16}$ + $Ba_4Mn_3O_{10}$ + $Ba_3Mn_2O_8$ in Figure 3b) appear over finite ranges of temperature and composition. In $Ba_3Mn_2O_8$, Mn is in the +5 oxidation state. Combined with the thermal analysis results, it can be inferred that in the temperature range of 550–650 °C, two types of reactions occur simultaneously:

(1) **Decomposition of $MnO_2$:**

$$4MnO_2 \rightarrow 2Mn_2O_3 + O_2\uparrow$$

(2) **Formation of high-valence manganate:**

$$3BaO + 2MnO_2 + \tfrac{1}{2}O_2 \rightarrow Ba_3Mn_2O_8$$

Although the second reaction consumes oxygen, the overall system still exhibits net oxygen loss under open conditions. With further increase in temperature, additional decomposition reactions occur, including the decomposition of $Mn_2O_3$ and $Ba_3Mn_2O_8$:

$$Ba_3Mn_2O_8 \rightarrow 2BaMnO_3 + BaO + \tfrac{1}{2}O_2\uparrow$$

Furthermore, $Ba_6Mn_5O_{16}$ and $Ba_4Mn_3O_{10}$ are derived from the decomposition of $Ba_3Mn_2O_8$ and subsequent transformations of $BaMnO_3$. These processes can be represented as:

$$5BaMnO_3 + BaO \rightarrow Ba_6Mn_5O_{16}$$

$$3BaMnO_3 + BaO \rightarrow Ba_4Mn_3O_{10}$$

$BaMnO_3$, $Ba_4Mn_3O_{10}$, and $Ba_6Mn_5O_{16}$ compete over broad multiphase coexistence regions in the phase diagram. To obtain high-purity single-phase products, the stoichiometric ratio and reaction temperature must be adjusted according to the corresponding single-phase fields in Figure 3b. The resulting reaction network couples valence evolution, oxygen exchange, and solid-state diffusion, in contrast to the sharp phase boundaries observed for Mg–V–O.

## 2.4 Two contrasting reaction regimes

The comparison between the Mg–V–O and Ba–Mn–O systems highlights two distinct regimes of solid-state reactivity under isothermal conditions. The Mg–V–O system evolves along a diffusion-controlled pathway with clean polymorphic transformations and no measurable oxygen exchange; once the reaction initiates, the temperature-dependent phase evolution is largely predictable. The Ba–Mn–O system, in contrast, involves continuous deoxygenation, manganese valence evolution from $Mn^{5+}$ in $Ba_3Mn_2O_8$ to $Mn^{4+}$ in the high-temperature barium manganates—superimposed on the $MnO_2 \rightarrow Mn_2O_3 \rightarrow Mn_3O_4$ decomposition sequence—and oxygen-dependent reaction equilibria. The result is a network of competing reaction channels that cannot be represented by a simple equilibrium phase diagram.

The two regimes require different synthesis strategies. For systems in the diffusion-controlled regime, thermodynamic predictions define the expected phase assemblages, while kinetic factors still set the boundary positions and the temperature dependence of the observed assemblages. For valence-coupled systems, the reaction atmosphere, oxygen partial pressure, and precursor selection become decisive variables, and reaction phase diagrams under explicit atmosphere conditions are required for reliable synthesis planning.

## 3. Discussion

**The finite-area three-phase fields in Figure 3b cannot be interpreted directly as equilibrium phase regions of a closed pseudo-binary section.** For a closed two-component system at fixed pressure, the condensed phase rule is $F = C - P + 1$; with $C = 2$ and $P = 3$, $F = 0$, so three-phase equilibrium occurs at an invariant temperature. A horizontal interval in a temperature–overall-composition projection may represent the range of overall compositions over which the same three phases coexist, but it is not a two-dimensional equilibrium three-phase field. The Ba–Mn–O samples additionally exchange oxygen with the furnace atmosphere and are measured after a finite reaction time, so their finite-area three-phase fields may reflect oxygen activity, reaction history, or kinetic coexistence rather than equilibrium phase topology alone. These fields therefore motivate measurements at controlled oxygen partial pressures and reaction times to determine how the observed boundaries depend on oxygen chemical potential and reaction history.

**The lowest-temperature appearance of $Ba_3Mn_2O_8$ shows that thermodynamic stability alone cannot predict the temperature-dependent phase evolution of a solid-state reaction.** $Ba_3Mn_2O_8$ is absent from the 0 K convex-hull reference, yet it is detected by ex situ XRD at approximately 550 °C, below the temperatures at which $BaMnO_3$, $Ba_6Mn_5O_{16}$, and $Ba_4Mn_3O_{10}$ are observed. This temperature-resolved observation indicates that kinetic accessibility can favor a phase not identified by the equilibrium energy ranking, but it does not establish the order in which phases form within an individual sample. $Ba_3Mn_2O_8$ may serve as a kinetically favored intermediate that facilitates subsequent transformations, although the present ex situ data cannot distinguish this possibility from a kinetically stabilized product or an equilibrium phase omitted from the reference dataset. Time-resolved and in situ measurements are required to establish its role in the reaction pathway.

**Reaction time should be treated as an explicit variable of a reaction phase diagram rather than as a deviation from an ideal equilibrium measurement.** The present diagrams represent phase assemblages obtained after the same 8 h reaction, which ensures direct comparability among temperature–composition points. Repeating the measurements at several reaction times would determine whether phase boundaries migrate, whether intermediate-phase regions contract or expand, and how long particular assemblages persist. The resulting time-dependent diagrams would provide experimental constraints on diffusion, nucleation, and oxygen-release processes that cannot be obtained from a single reaction time.

**The formation of a phase at high temperature does not guarantee its retention at room temperature.** $\gamma$-$Mg_2V_2O_7$ forms near 911 °C but is not recovered after cooling, so the ex situ XRD maps identify the assemblages retained after the complete thermal cycle rather than every phase present at the reaction temperature. In situ high-temperature diffraction, combined with controlled quenching, would allow the formation window and the room-temperature recovery window to be determined separately. For metastable oxides, these two windows must both be established to define a usable synthesis route.

## 4. Conclusion

We have shown that an automated laboratory can construct reaction phase diagrams under a fixed 8 h reaction time by combining parallel synthesis, XRD characterization, and iterative refinement of phase boundaries. The $MgO$–$V_2O_5$ and $BaO$–$MnO_2$ sections exhibit contrasting behavior: the former reacts without measurable mass loss and develops well-defined diffusion-controlled boundaries, whereas the latter undergoes continuous oxygen loss and forms extended multiphase regions. Reaction phase diagrams thus provide experimental maps of phase accessibility under defined synthesis conditions.

Measurements as a function of reaction time, oxygen partial pressure, and cooling pathway, supported by in situ characterization, would connect these maps to reaction mechanisms and phase retention. A standardized database that records precursors, thermal history, atmosphere, and cooling conditions together with the observed phase assemblages would complement equilibrium phase diagrams and provide directly comparable data for synthesis planning.

## Acknowledgements

This research is supported by National Key R&D Program of China (Grant No. 2024YFF0508500), Chinese Academy of Sciences (Grant No. YSBR047).

## Author contributions

M.L. conceived and supervised the project. J.H. performed the experiments, analyzed the data, and drafted the manuscript. J.F. and Y. H. contributed to the construction and development of the automated laboratory platform. S.M. and W.W. contributed to discussion and revision of the manuscript. All authors discussed the results and approved the final version.

## Competing interests

The authors declare no competing financial or personal interests that could have appeared to influence the work reported in this paper.